# Hydrogen in layered iron arsenide: indirect electron doping to induce superconductivity

Taku Hanna,[1] YoshinoriMuraba,[1] Satoru Matsuishi,[1] Naoki Igawa,[2] Katsuaki Kodama,[2] Shin-ichi Shamoto[2] and Hideo Hosono[1,3,*]

[1]*Materials and Structures Laboratory, Tokyo Institute of Technology, 4259 Nagatsuta-cho, Midori-ku, Yokohama 226-8503, Japan*

[2]*Quantum Beam Science Directorate, Japan Atomic Energy Agency, Tokai, Ibaraki 319-1195, Japan.*

[3]*Frontier Research Center, Tokyo Institute of Technology, 4259 Nagatsuta-cho, Midori-ku, Yokohama 226-8503, Japan*

Utilizing the high stability of calcium and rare earth hydrides, $CaFeAsF_{1-x}H_x$ ($x$ = 0.0-1.0) and $SmFeAsO_{1-x}H_x$ ($x$ = 0.0-0.47) have been first synthesized using high pressure to form hydrogen-substituted 1111 type iron-arsenide superconductors. Neutron diffraction and density functional calculations have demonstrated that the hydrogens are incorporated as $H^-$ ions occupying $F^-$ sites in the blocking layer of CaFeAsF. The resulting $CaFeAsF_{1-x}H_x$ is non-superconducting, whereas $SmFeAsO_{1-x}H_x$ is a superconductor, with an optimal $T_c$ = 55 K at $x \sim$ 0.2. It was found that up to 40% of the $O^{2-}$ ions can be replaced by $H^-$ ions, with electrons being supplied into the FeAs-layer to maintain neutrality ($O^{2-} = H^- + e^-$). When $x$ exceeded 0.2, $T_c$ was reduced corresponding to an electron over-doped region.

*Corresponding author: Hideo Hosono
Materials and Structures Laboratory, Tokyo Institute of Technology
4259 Nagatsuta, Midori, Yokohama 226-8503, Japan
TEL +81-45-924-5359
FAX +81-45-924-5339
E-mail: hosono@msl.titech.ac.jp

Since the discovery of superconductivity in LaFeAsO$_{1-x}$F$_x$ ($T_c$= 28 K),[1] layered iron pnictides and related materials have been intensively investigated as candidates for high-$T_c$ superconductors. To date, various types of iron-based superconductors have been synthesized[2-11], with the highest $T_c$ of ~ 55 K being recorded in SmFeAsO$_{0.9}$F$_{0.1}$.[12-17] The 1111type iron arsenides $Ln$FeAsO with a ZrCuSiAs-type structure[18] are composed of stacks of alternating FeAs anti-fluorite-type conducting layers and $Ln$O fluorite-type blocking layers. Although the parent compounds are non-superconducting at ambient pressures, they are superconducting when electrons are doped into the FeAs-layer via fluorine-substitution at the oxygen sites (O$^{2-}$ = F$^-$ + $e^-$),[1,13-17] oxygen vacancy formation (O$^{2-}$ = V$_O$ + 2$e^-$),[19-21] or transition metal (a $TM$ with excess electrons when compared with Fe, like Co or Ni) substitution into the iron site.[22-25] $TM$-substitution, adding electrons to the FeAs-layer, is a "direct" doping mode, whereas F-substitution or O-vacancy formation is an "indirect" mode, because impurities are doped to the blocking layers and the electrons transferred to the FeAs layers. The general consensus is that indirect doping gives a higher$T_c$, because less structural and electrical disturbances occur in the superconducting layers. However, over-doping via the indirect mode has not been confirmed for most 1111type iron-arsenides. In other words, further improvement in the $T_c$ may be possible. For example, the solubility limit of fluorine substitution in $Ln$FeAsO is lower than 20%. This fact is in contrast to alkali metal substitution in 122 type Ba$_{1-x}$K$_x$Fe$_2$As$_2$ superconductors, in which the Ba site can be fully replaced by K.[26] This restraint also makes it difficult to complete the electronic phase diagram (doping level vs. critical temperatures) beyond the optimal doping level. Completion of the phase diagram is important, not only to improve $T_c$, but also to understand superconductivity. Besides, although several papers have reported on the variation of $T_c$ with nominal oxygen vacancy or hydrogen content in the 1111 system, the chemical states and the analyzed composition have not been clarified to date.[19,20,27-29] This ambiguity is an obstacle to understanding the emerging and carrier concentration dependence on $T_c$.

In this paper, we propose a novel method for indirect electron doping to induce superconductivity in 1111type iron arsenides. By using a high pressure synthesis method with an excess hydrogen source, CaFeAsH has been synthesized, a new analog of $Ln$FeAsO, and the solid solution between CaFeAsF[30,31] and CaFeAsH (CaFeAsF$_{1-x}$H$_x$ where 0 <$x$≤ 1). While CaFeAsF$_{1-x}$H$_x$ is non-superconducting like $Ln$FeAsO, these results indicate the anion site in the blocking layer can be substituted by H$^-$. Considering the facile formation of metal hydrides with rare earths, H$^-$ appears to be stabilized in these rare earth compounds. Thus, the hydrogen substitution technique has been applied to electron doping in SmFeAsO as with fluorine substitution (O$^{2-}$ = H$^-$ + $e^-$). Up to 40 % of the oxygen sites were successfully replaced by hydrogen (SmFeAsO$_{1-x}$H$_x$, 0 <$x$≤ 0.4) and superconductivity with a maximum $T_c$ = 55 K ($x$ ~ 0.2) was induced. Above 20% hydrogen substitution, $T_c$ was decreased, indicating that electron over-doping, which was not seen with F substitution in this system, was observed.[32-34]

CaFeAsH was synthesized by the solid state reaction of CaAs, $Fe_2As$ and $CaH_2$ with $LiAlH_4$ as an excess hydrogen source, at 1000°C and 2 GPa (CaAs + $Fe_2As$ + $CaH_2 \rightarrow$ 2CaFeAsH). A belt-type anvil cell was employed for the high pressure synthesis. Powders of CaAs and $Fe_2As$ were prepared from their respective metals (Ca: 99.99% Aldrich, Fe: 99.9% Kojyundo, As: 99.9999% Kojyundo) and $CaH_2/CaD_2$ were synthesized by heating calcium metal in a $H_2/D_2$ atmosphere. All starting materials and precursors for the synthesis were prepared in a glove-box (Miwa Mfg.) filled with purified Ar gas ($H_2O$, $O_2$ < 1 ppm). The mixture of starting materials was placed into a BN capsule. Following the internal hydrogen source technique developed by Fukai and Okuma,[35,36] $LiAlH_4$ (98%, Tokyo Kasei) was also placed in the capsule, with a BN separator, as a supplementary hydrogen source. The deuterated analog, CaFeAsD, was also prepared using $CaD_2$ and $LiAlD_4$ (98%, Aldrich). The resulting crystalline phases were identified by powder X-ray diffraction (XRD). Crystalline phases in the resulting samples were identified by powder X-ray diffraction (XRD) using a Bruker diffractometer model D8 ADVANCE (Cu rotating anode). The Rietveld analysis of XRD patterns was performed using TOPAS code.[37] Figure 1(a) shows powder XRD patterns of the resulting products at room temperature (RT). Except for several minor peaks due to metal Fe and $Ca(OH)_2/Ca(OD)_2$ phases (<5 wt.%), all the peaks could be indexed to a tetragonal ZrCuSiAstype[18] structure (space group: $P/4nmm$) with lattice constants $a$ = 0.3878 nm and $c$ = 0.8260 nm for CaFeAsH and $a$ = 0.3876 nm and $c$ = 0.8257 nm for CaFeAsD. The differences in the lattice constants between the hydride and deuteride versions were less than 0.05%. Rietveld analysis indicated that the observed XRD pattern is well explained by assuming a model structure composed of alternate layers of FeAs and CaH(D). However the site position and occupancy of hydrogen cannot be determined by XRD, because the X-ray atomic scattering factor of $H^-/D^-$ is too small.

The amount of hydrogen incorporated in the resulting samples was evaluated by thermogravimetry/mass spectroscopy (TG-MS) performed using a Bruker AXS TG-DTA/MS9610 equipped with a gas feed port to inject the standard $H_2$ gas into the sample chamber. 20 mg of sample was heated up to 800 °C with a heating rate of 20 K·min$^{-1}$ under a helium gas flow. Hydrogen released from sample, in the form of $H_2$ molecule was ionized and detected by a quadrupole mass spectrometer as an ion with $m/z$ = 2. As shown in Fig. 1 (b), weight loss involving the emission of $H_2$ molecule was observed from 200 to 600°C. The TG-MS measurement continued to 800°C, where the sample decomposed into a mixture of $CaFe_2As_2$, FeAs and unknown phases probably consisting of Ca with $O_2$, $N_2$ and/or $H_2O$ gas from the He gas flow. The amount of released $H_2$ was estimated to be 3.08mmol/g from the integration of the mass peak $m/z$ = 2. This quantity was almost equal to that expected for the decomposition of CaFeAsH (2CaFeAsH →Ca + $CaFe_2As_2$ + $H_2\uparrow$, 2.91mmol/g).

Since the coherent neutron scattering cross section of deuterium (5.592 barn) is comparable to that of Ca (2.78), Fe (11.22) or As (5.44),[38] the atomic position and site occupancy of deuterium can be determined by neutron powder diffraction (NPD). Neutron powder diffraction of CaFeAsD was measured on ~3 grams of sample by the high resolution powder diffractometer (HRPD) installed at the JRR-3 reactor of JAEA (beam collimation of 30'-40'-(sample)-6' with neutron wavelength λ = 0.182391 nm). Figure 1(c) shows NPD patterns observed at RT and 10 K. Rietveld analyses performed using RIETAN-FP code[39] revealed that the anion site in the block layer is occupied by deuterium, with the site occupancy of 0.935. Taking into account the isotopic purity of the LiAlD$_4$ (98 %) and inclusion of hydrogen from other starting materials, we conclude that the remaining fraction (0.065) of the anion site is primarily occupied by hydrogen. The NPD pattern observed at 10 K was attributed to an orthorhombic phase, space group: *Cmma*, *a*=0.549213(16) nm, *b*=0.545660(16) nm, *c*=0.821154(24) nm. This crystal symmetry change is the same as in other 1111type iron arsenides (*Ln*FeAsO and *Ae*FeAsF) with tetragonal to orthorhombic transitions reported in the range of 120-180 K. High pressure synthesis appears to be essential for the formation of CaFeAsH, because no such a phase could be prepared at ambient pressures.

Figure 2 (a) shows the temperature dependence of dc electrical resistivity (ρ) in CaFeAsH and CaFeAsD measured from 300 to 2 K, using a physical properties measurements system (PPMS, Quantum Design). Both samples exhibit sudden decreases at ~100 K ($T_{anom}$). This anomaly has also been observed in 1111 and 122 type iron arsenides and has been attributed to a tetragonal to orthorhombic transition.

Figure 2 (b) shows the calculated total and projected density of states (DOS and PDOS) of CaFeAsH obtained by density functional calculations using density functional theory (DFT), the generalized gradient approximation functional PBE,[40] and the projected augmented plane waves method [42] implemented in the Vienna ab initio simulation program (VASP) code,[42] following refs. [43,44] and 50. A √2a×√2b×2c supercell containing 8 chemical formulae was used for the calculation and the plane-wave basis set cutoff was set to 600 eV. For Brillouin zone integrations to calculate the total energy and DOS, 4×4×2 Monkhorst-Pack grids of *k*-points were used. The total energy was minimized with respect to both the coordinates of all atoms and lattice parameters. Then, the stripe type anti-ferromagnetic ordering of the Fe spins, commonly observed for 1111 type Fe arsenides below $T_{anom}$, was obtained as the most stable magnetic structure. To obtain the projected DOS, the charge density was decomposed over the atom-centered spherical harmonics with Wigner-Seitz radius $r = (3V_{cell}/4\pi N)^{1/3}$, where $V_{cell}$ and $N$ are the unit cell volume and the number of atoms in an unit cell, respectively. Total DOS and Fe PDOS profiles revealed the semi-metallic nature of CaFeAsH, with Fe 3d up-spin and down-spin bands overlapped at the Fermi level ($E = 0$). Hydrogen 1s levels were located at ~2 eV below the Fermi level and were fully occupied. It is evident from these results that hydrogen is incorporated as H$^-$ (1s$^2$) in CaFeAsH. This finding is consistent with the similarities in the ρ-*T* curves of CaFeAsH and CaFeAsF, indicating that the replacement of F sites with H$^-$ ions does not seriously affect the electronic structure of theFeAs conduction layer.

Based on the similarity of the charge and size of the hydride and fluorine ion, the formation of the solid solution CaFeAsF$_{1-x}$H$_x$ was expected, and was obtained over the full *x* range by adding CaF$_2$ to the starting mixture in the high pressure synthesis, CaAs + Fe$_2$As + (1−*x*) CaF$_2$ + *x* CaH$_2$ →

2CaFeAsF$_{1-x}$H$_x$. Figure 2 (c) shows the lattice parameters $a$ and $c$ as a function of $x$. The value of $x$ in the resulting CaFeAsF$_{1-x}$H$_x$ was evaluated by TG-MS measurement. While the lattice parameter $a$ was almost independent of $x$, the value of $c$ was proportional to $x$, indicating that the geometry of the CaF$_{1-x}$H$_x$ layer is determined by the weighted average of the ionic radius of F$^-$ ($r_F$) and H$^-$ ($r_H$). From the F-Ca distance ($r_F + r_{Ca}$ = 233.7 pm)[45] and the H-Ca bond length ($r_H + r_{Ca}$ = 230.2 pm) evaluated from the Rietveld analysis, the ionic radius of H$^-$ was estimated for 1111type iron-arsenides. On the assumption that a F$^-$ coordinated by four Ca$^{2+}$ ions retains Shanon's ionic radius ($r_F$ = 131 pm),[34] then the ionic radius of Ca$^{2+}$ was calculated to be 102.7 pm. The radius of H$^-$ calculated from the H-Ca bond length(230.2 pm) in CaFeAsH is 127.5 pm, which is smaller than that of F$^-$ by 2.8%. Chevalier et al.[46] reported the synthesis of 1111type hydrides $LnMX$H ($Ln$ = lanthanides; $M$ = Mn, Fe and Co; $X$ = Si and Ge). These crystals with stacks of alternating $LnH$ hydride layers and $MX$ layers were made by hydrogen insertion to CeFeSi-type [47]$LnMX$. In contrast, the present high pressure synthesis using hydrides does not require the precursor.

Next, the hydrogen substitution technique was applied to electron doping in SmFeAsO. Like CaFeAsF$_{1-x}$H$_x$, SmFeAsO$_{1-x}$H$_x$ was prepared by the solid state reactions of SmAs, FeAs, Fe$_2$As, Sm$_2$O$_3$ and SmH$_2$ at 2 GPa at 1200 °C; (2+$x$)SmAs + (2+$x$) Fe$_2$As + (2−2$x$)FeAs + (2−2$x$) Sm$_2$O$_3$ + 3$x$ SmH$_2$→6SmFeAsO$_{1-x}$H$_x$. SmAs and SmH$_2$ were prepared by the reaction of Sm metal with arsenic or hydrogen gas. In this case, a mixture of NaBH$_4$ and Ca(OH)$_2$ was used as the excess hydrogen source. Figure 3 (a) shows representative powder XRD patterns of SmFeAsO$_{1-x}$H$_x$(nominal $x$ values in starting mixtures are 0.1, 0.3 and 0.5). When nominal $x$ exceeds ~0.4, the SmAs phase begins to segregate, implying the actual hydrogen content ($x$) is lower than the nominal $x$. Figure 3(b) shows TG-MS profile ($m/z$ = 2) of SmFeAsO$_{1-x}$H$_x$ with nominal $x$ = 0.15. Hydrogen incorporated in SmFeAsO$_{1-x}$H$_x$ was released between 400 and 800ºC. The shape of the emission curve is close to that of CaFeAsF$_{1-x}$H$_x$. Figure 3 (c) compares the hydrogen content ($x$) and the deficient amount of oxygen ($y$) in prepared samples per chemical formula (SmFeAsO$_{1-y}$H$_x$), as a function of nominal $x$ in the starting mixture. The former value was determined by TG-MS and the latter was measured with a JEOL JXA-8530F electron-probe micro-analyzer (EPMA) equipped with a field emission-type electron gun and wavelength dispersive X-ray (WDX) detectors. The micrometer scale compositions within the main phase were probed on 5-10 focal points and the results were averaged. For nominal $x \leq 0.4$, the hydrogen content agrees with $y$ and the nominal $x$, indicating the oxygen site (O$^{2-}$) was successfully substituted with hydrogen (H$^-$). Figure 3 (d) shows the variation of lattice constants $a$ and $c$ of SmFeAsO$_{1-x}$H$_x$ system. Same as CaFeAsF$_{1-x}$H$_x$, the decrease of $c$ with $x$ corresponds to the difference in ionic radii between O$^{2-}$ (142 pm) and H$^-$ (127.5 pm in CaFeAsF$_{1-x}$H$_x$). On the other hand, $a$ of the SmFeAsO$_{1-x}$H$_x$ system apparently decreases with $x$. This result contrasts with the CaFeAsF$_{1-x}$H$_x$ system, in which $a$ is independent of $x$. It is likely to be a result of electron doping in the FeAs layer by H substitution (O$^{2-}$ = H$^-$ + e$^-$), i.e., the doped electrons occupy the bonding states (i.e., a hole pocket) of FeAs layer, and shorten the intra-layer Fe-Fe distance.

Figure 4(a) shows the temperature dependence of the electrical resistivity in SmFeAsO$_{1-x}$H$_x$. A sudden drop of resistivity due to superconductivity was observed for $x \geq 0.03$ and the maximum critical temperature $T_c$ (onset) was 55 K at $x \sim 0.2$. As $x$ increased over 0.2, $T_c$ decreased, indicating the appearance of an over-doped region. To reveal the change in transport properties of normal conducting state with hydrogen substitution, we examined the exponent $n$ of $\rho(T) \sim T^n$ for each sample by fitting the data to $\rho(T) = \rho_0 + AT^n$ in the temperature range between $T$ just above $T_c$ and $T = 130$ K, where $\rho_0$ is the residual component of resistivity. Figure 4(b) shows the results; $n$ decreases from 2 to 1 with $x$ below the optimal substitution level ($x < 0.2$), while $\rho_0$ rapidly decreases to 0. The $T^2$ dependence up to high temperature is characteristic for strongly correlated Fermi liquid systems, in which electron–electron interaction dominates the scattering of the carriers.[48] $T$ linear dependence at the optimal level indicates non-Fermi liquid state was induced by electron doping via hydrogen substitution. Similar behaviors have been also observed in 122 type and some 1111 type iron arsenides.[49,50] On the contrary, $n$ deceases to 0.5 in the over-doping region ($x > 0.2$). This is quite different from 122 type arsenide in which $n$ is again close to ~2 in the over doping region. Figure 4(c) shows magnetic susceptibility ($4\pi\chi$) vs. $T$ plots of $x = 0.03$-0.47 samples under zero-field cooling and field-cooling (FC) with a magnetic field of 10 Oe. The diamagnetism due to superconductivity was observed below $T_c$ (mag) for $x = 0.07$-0.43. Shielding volume fraction (SVF) of each sample was obtained from the gradient of the liner region in Magnetization (M) vs. Field (H) plot observed at 2 K (Fig. 4(d)). The SVF above 45 % observed for $x = 0.07$-0.43 indicates the bulk superconductivity, whereas the SVF in $x = 0.03$ and 0.47 samples is lower than 1%. Such a low SVF implies that the hydrogen concentration is inhomogeneous in these samples and only the specific region with preferred hydrogen substitution levels shows superconductivity. The sample of $x = 0.03$ shows both the sudden drop of resistivity due to superconducting transition ($T_c = 5$ K) and the $\rho$-T anomaly ($T_{anorm} = 114$ K) due to crystallographic transition accompanied with magnetic ordering of the Fe spins. Similar behavior has been observed in SmFeAsO$_{1-x}$F$_x$ around $x \sim 0.04$.[32] These data imply the coexistence of superconductivity and magnetic ordering in the limited $x$ range near 0.03-0.04. While such coexistence has been confirmed in 122 type iron arsenide superconductors,[26] it is hard to distinguish from the coexistence in systems with compositional disorder. Detailed experiments using an appropriate microprobe such as μSR are required to clarify the intrinsic nature of this coexistence. Figure 4 (e) shows $x$-$T$ diagram of SmFeAsO$_{1-x}$H$_x$ superimposed with that of SmFeAsO$_{1-x}$F$_x$ with the fluorine content $x$ measured by EPMA as reported by A.Köhler et al.[33] $T_c$ vs. $x$ plots of SmFeAsO$_{1-x}$H$_x$ and SmFeAsO$_{1-x}$F$_x$ overlap at $x < 0.15$, indicating that hydrogen gives indirect electron doping to the FeAs-layer just like fluorine. While the solubility limit of fluorine in the oxygen site is restricted to less than 20 % ($x = 0.2$),[33,34] hydrogen can reach 40 %. The wider substitution range is useful for the optimization of the electron doping level to induce superconductivity and to complete the electronic phase diagram, including the over-doped region.

In summary, 1111type CaFeAsF$_{1-x}$H$_x$ ($0 < x \leq 1$) and SmFeAsO$_{1-x}$H$_x$ ($0 < x \leq 0.4$) were first synthesized by a high pressure technique with an excess hydrogen source. Substitution of $H^-$ into the $F^-$ site in CaFeAsF was confirmed by neutron powder diffraction analysis and density functional calculations. The resulting $CaFeAsF_{1-x}H_x$ were non-superconducting, with a $\rho$-$T$ anomaly at ~100 K due to a tetragonal to orthorhombic transition commonly observed in $Ae$FeAsF and $Ln$FeAsO. In $SmFeAsO_{1-x}H_x$ ($0 < x \leq 0.4$), $H^-$ substitution into the $O^{2-}$ site supplies electrons to the FeAs-layer, giving raise to superconductivity with an optimized $T_c$ of 55 K. Further $H^-$-substitution beyond the optimal level leads to a decrease in $T_c$ up to $x= 0.4$. The present results suggest the hydrogen ($H^-$) substitution to the anion site is generally applicable for electron doping in 1111type iron-arsenide superconductors.

We thank Toshiyuki Atou and Osamu Fukunaga of the Tokyo Institute of Technology for their valuable discussions. This research is funded by the Japan Society for the Promotion of Science (JSPS) through the FIRST program, initiated by the CSTP.

**Figure captions**

Fig.1 (a) Powder XRD patterns of CaFeAsH and CaFeAsD at room temperature. (b) TG and MS ($m/z = 2$ corresponds to the $H_2$ molecule) profiles of CaFeAsH. Weight loss due the decomposition of sample with hydrogen emission was observed from 200 to 600°C. Hydrogen concentration was estimated to be 1.05 molecules per unit cell by integration of the MS trace curve. (c) Neutron powder diffraction pattern of CaFeAsD observed at 300 K and 10 K (d) Crystal structure of CaFeAsD(H). (d) Crystal structure of CaFeAsH at 300 K (tetragonal) and 10 K (orthorhombic)

Fig.2 (a) ρ-$T$ profile of CaFeAsH and CaFeAsD compared with CaFeAsF. (b) Total and atomic projected density of states of CaFeAsH obtained by density functional calculation using VASP code. The origin was set to the Fermi Level. (c) Variation of lattice constants ($a, c$) as function of $x$ in the solid solution CaFeAsF$_{1-x}$H$_x$. Hydrogen content ($x$) was estimated by TG-MS method.

Fig.3 Structural and chemical composition data on SmFeAsO$_{1-x}$H$_x$. (a) Powder XRD patterns of SmFeAsO$_{1-x}$H$_x$ (nominal $x$ = 0.3, 0.4 and 0.5). Insets show close-up views. (b) TG and MS ($m/z = 2$) profiles with nominal $x$ = 0.15. $H_2$ gas emission associated with weight loss was observed from 400 to 800°C. (c) Oxygen deficiency content determined by EPMA measurement ($y$) and hydrogen content estimated by TG-MS method ($x$) in SmFeAsO$_{1-y}$H$_x$ as a function of nominal $x$ in starting mixture. The measured $x$ is almost equal to $y$ and nominal $x$, indicating the deficiency of oxygen site is wholly compensated by occupation of hydrogen. (d) Lattice constants ($a$ and $c$) as a function of $x$.

Fig.4. Electrical and magnetic properties. (a) ρ-$T$ profiles of SmFeAsO$_{1-x}$H$_x$ in under-doped (left; $x$ = 0.0-0.19)and over-doped states (right; $x$ = 0.22-0.47). (b) log (ρ($T$) − ρ$_0$) vs. log$T$ plots for data on non-superconducting region, where ρ$_0$ is the residual component of resistivity obtained from fitting (see text). Lowers show ρ$_0$ and the exponent residual resistivity $n$ as a function of $x$. (c) Zero-field cooling (ZFC) and field cooling (ZFC) 4πχ-$T$ curves measured under the magnetic field ($H$) of 10 Oe (d) Magnetization ($M$) vs. $H$ curves and shielding volume fraction vs. $x$ plot at 2 K. (e) $x$-$T$ diagram of SmFeAsO$_{1-x}$H$_x$ superimposed by that of SmFeAsO$_{1-x}$F$_x$.[33]

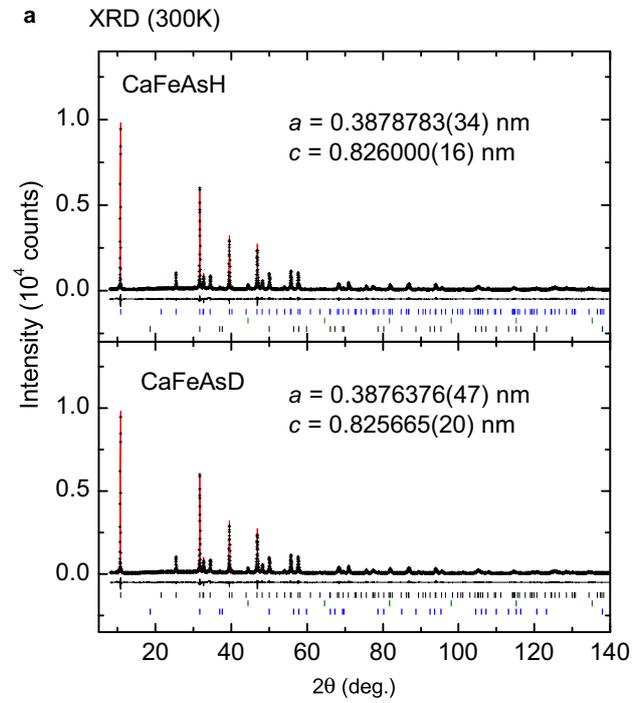

**a** XRD (300K)

CaFeAsH
$a = 0.3878783(34)$ nm
$c = 0.826000(16)$ nm

CaFeAsD
$a = 0.3876376(47)$ nm
$c = 0.825665(20)$ nm

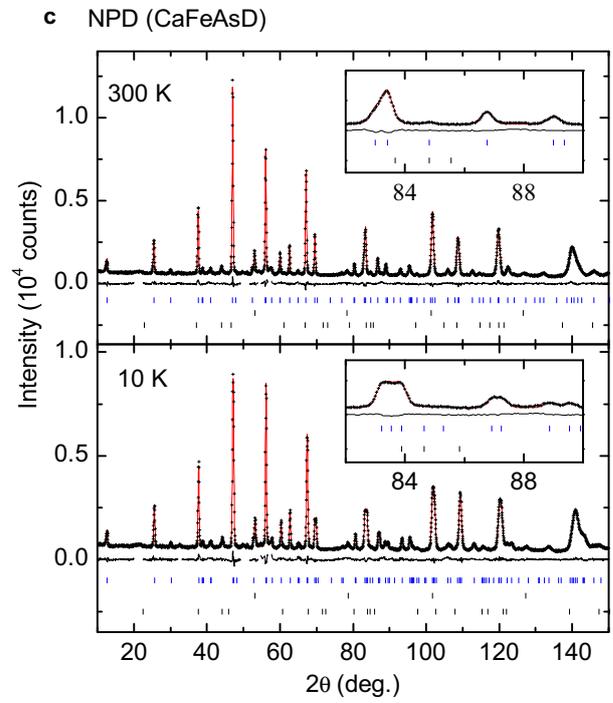

**c** NPD (CaFeAsD)

300 K

10 K

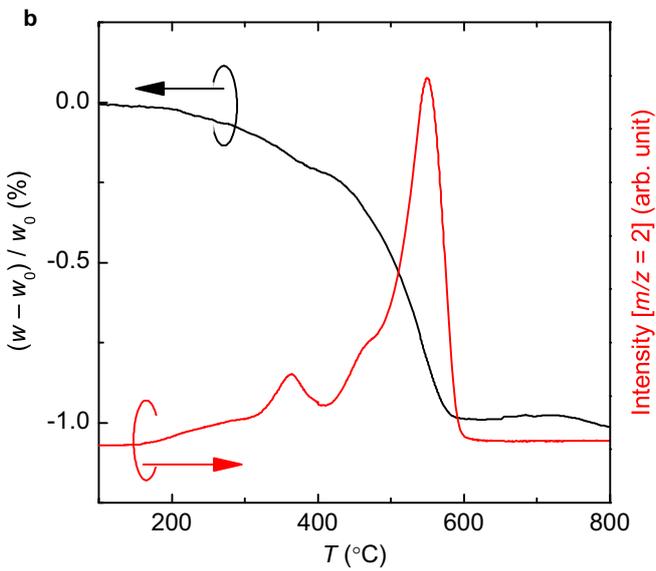

**b**

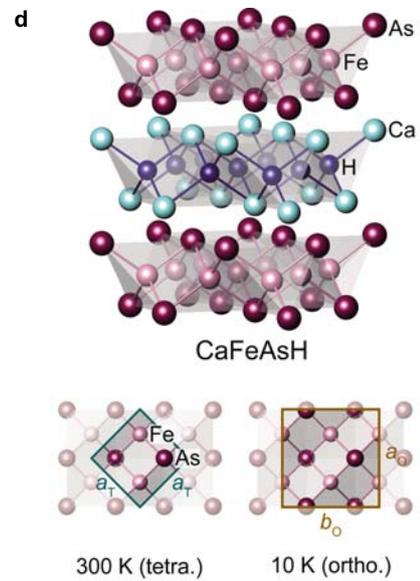

**d**

CaFeAsH

300 K (tetra.)   10 K (ortho.)

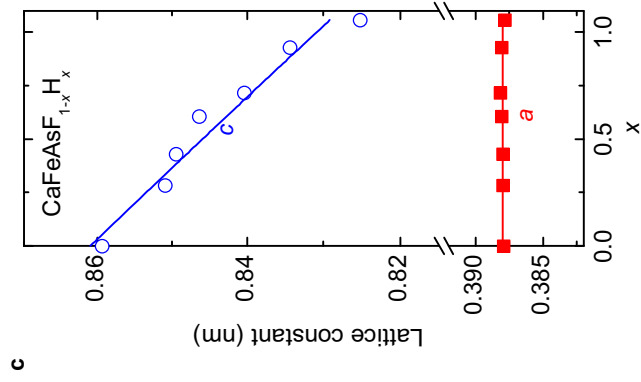

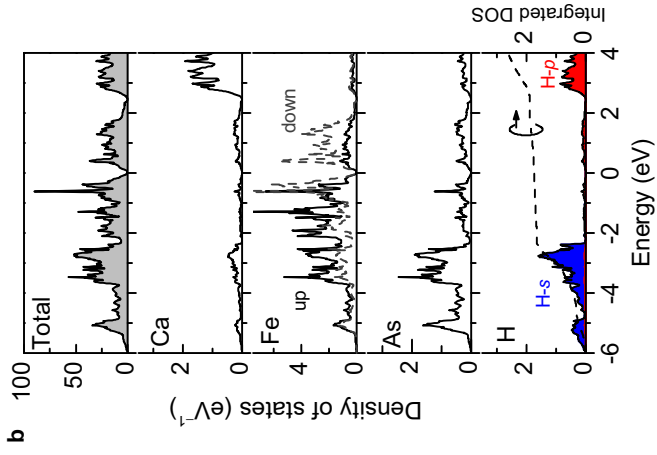

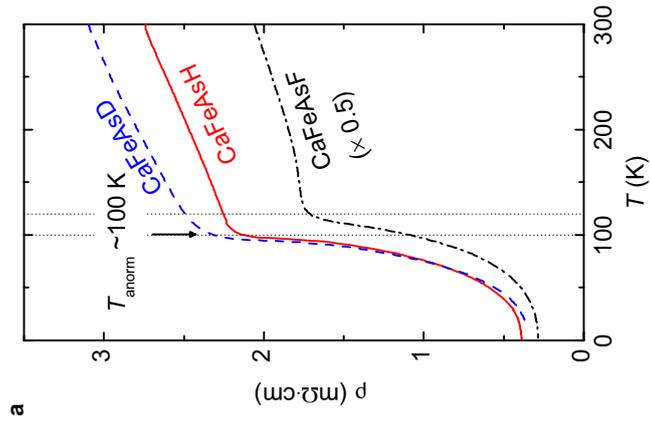

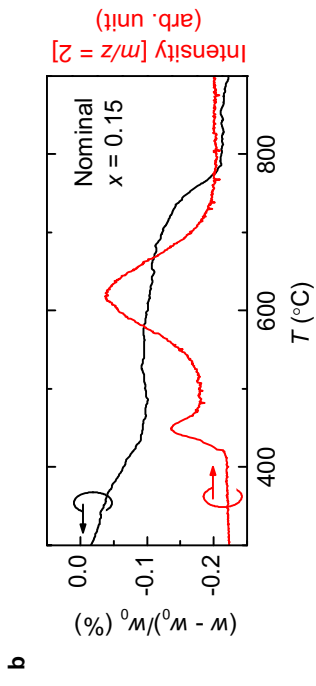
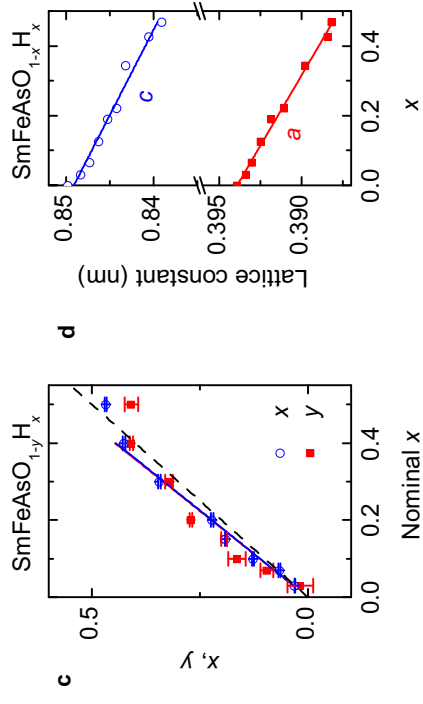
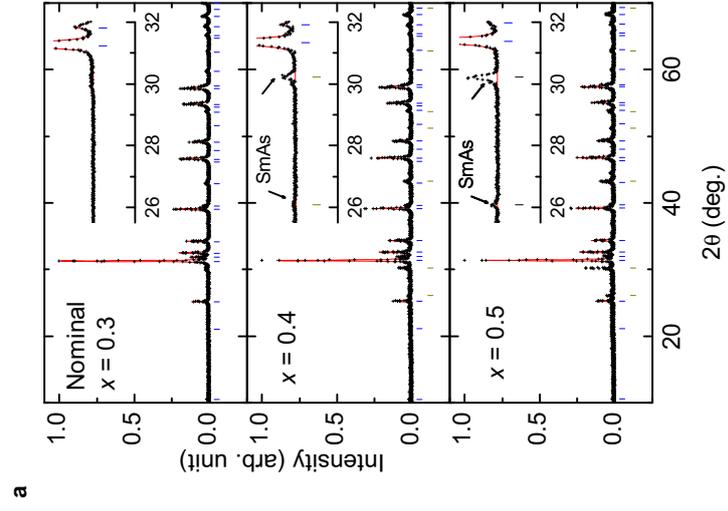

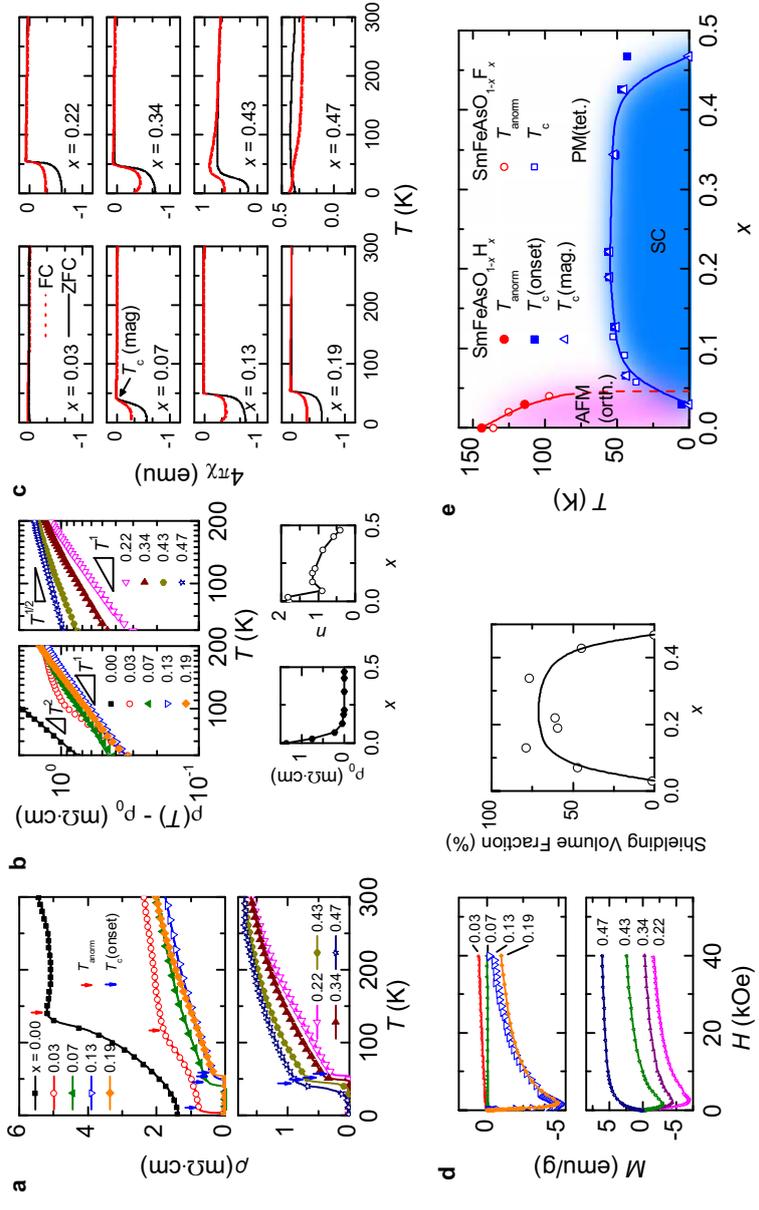